\documentclass[%
 reprint,
showpacs,preprintnumbers,
 amsmath,amssymb,
 aps,
prc,
twocolumn,
superscriptaddress]{revtex4-2}
\usepackage[english]{babel}
\usepackage[utf8]{inputenc} 
\usepackage[T1]{fontenc}
\usepackage{graphicx}
\usepackage{dcolumn}
\usepackage{bm}
\usepackage{float}

 \usepackage{amsthm}
\usepackage{mathtools}
\usepackage{xcolor} 
\usepackage{verbatim}
\usepackage[colorlinks,citecolor=blue,linkcolor=blue,anchorcolor=blue,filecolor=blue,urlcolor=blue]{
hyperref} 
\begin{document}

\title{Inclusive Breakup Cross Sections in Reactions Induced by the Nuclides \textsuperscript{6}He and \textsuperscript{6,7}Li in the Two-Body Cluster Model} 

\author{L.~A.~Souza}
\affiliation{
\mbox{Departamento de F\'isica, Instituto Tecnol\'ogico de Aeron\'autica, DCTA, 
12228-900, S\~ao Jos\'e dos Campos, SP, Brazil}
}

\author{E.~V.~Chimanski}
\affiliation{
\mbox{Lawrence Livermore National Laboratory    
7000 East Avenue, Livermore, CA 94550, U.S.A }
}

\author{B.~V.~Carlson}
\affiliation{
\mbox{Departamento de F\'isica, Instituto Tecnol\'ogico de Aeron\'autica, DCTA, 
12228-900, S\~ao Jos\'e dos Campos, SP, Brazil}
}
\date{\today}

\begin{abstract}
  Calculations of inclusive $\alpha$ particle production are performed to interpret reaction experimental data induced by the two-neutron halo $^6$He and the isotopes $^{6,7}$Li on different target nuclei. 
We have implemented the zero-range post-form DWBA to compute the elastic and nonelastic breakup cross sections. Integrated cross sections, angular distributions and spectra are presented. 
Projectiles are approximated as two-body clusters,  namely $^6$He as $\alpha$+dineutron and
$^6$Li ($^7$Li) are interpreted as $\alpha$+deuteron(triton).
The S\~ao Paulo optical potential is employed in the calculation of the distorted wave functions for the incident channels, while standard phenomenological interactions are taken for the fragment-target interactions.
%
Calculations for $^6$Li and $^7$Li  furnish a good description of the data, while the reaction involving $^6$He is found to be more complicated, where interpretation of the data must still be considered incomplete. The present analysis identifies an overall large contribution from inclusive breakup emission for all the cases studied.

\end{abstract}

\maketitle

\section{Introduction}

The breakup of weakly bound nuclei, such as the isotopes $^6$He, $^6$Li and $^7$Li, leads to the production of light fragments, e.g. nucleons, deuterons and $\alpha$ particles as the result of projectile dissociation.
With respect to $^6$He, its unusual structure, consisting of a two-neutron halo extending far from the $\alpha$ core, has attracted the attention of researchers in the last decades. Its halo property can be studied using nuclear reaction experiments providing information about the distribution of matter of nuclei far from stability. $^6$He has a short half-life of 0.8 seconds, which leads to experimental difficulties in producing this nucleus to perform measurements. Nevertheless, a number of groups have dedicated their time and efforts to understand the reaction mechanisms involved in the $^6$He induced events ~\cite{sanchez-benitez2008, dipietro2004,appannababu2019,signorini2001,defaria2010,benjamin2007}. The extended spatial distribution of this weakly bound nucleus points to a competition between partial fusion (transfer) and breakup of the projectile. These two mechanisms have cross sections larger than the contribution from the total fusion component at energies near the Coulomb barrier. 
In Ref.~\cite{escrig2007} it was found that the reaction cross section of $^6$He +$^{208}$Pb at backward angles is dominated by a prominent $\alpha$ group. Such emissions were interpreted as direct transfer of the halo neutrons to the target. Fusion and transfer + breakup channels on the medium-mass target $^{64}$Zn have been studied by Di Pietro {\it et al.}~\cite{dipietro2004}. They reported a strong yield of $\alpha$ particles coming from both transfer and breakup processes, while no effect on the fusion cross section at energies above the Coulomb barrier was observed.
 
One of the challenges regarding two neutron (2n) emission is associated with their spatial correlation and their connection to a dineutron-like model for the two-neutron halo. A. B. Migdal \cite{migdal} pioneered such a picture by showing that the 2n state may be interpreted as a dineutron coupled to a core in a three-body interaction model. His approach works because the force required to bind two neutrons is sufficiently weak to be unable to bind a single neutron to the core. More recently, several groups have used similar ideas in the investigation of two-neutron transfer and Coulomb breakup \cite{Zhukov1993,hagino2007,nakamura2006,Jensen2004}. Evidence for a 2n correlation leading to a dineutron structure was first observed in the decay of $^{16}$Be  in \cite{Spyrou2012} and also for  $^{11}$Li, in two recent works by Kubota {\it et al.}~\cite{kubota2020} and by Sun {\it et al.}~\cite{sun2021}. In these works, the average anglular correlation between the two neutrons is measured and is found to be smaller than the uncorrelated angle  (90$^\circ$). Moreover, Kubota {\it et al.}~\cite{kubota2020}, show that the average anglular distribution is asymmetric while the momentum dependence indicates that the dineutron correlation is localized radially on the $^{11}$Li surface.


Many of the results obtained for $^6$He can be compared to those of the less problematic lithium isotopes. These nuclei have the advantage of being stable and breaking up into well-known fragments. For example, a reaction induced by $^6$Li often results in fragmentation into $\alpha$+deuteron and while one induced by $^7$Li yields $\alpha$+triton.
A larger probability of $\alpha$ production in comparison to other fragments in reactions of $^6$Li+$^{208}$Pb, is suggested in Ref.~\cite{signorini2004}. For the $^7$Li case in Ref.~\cite{badran2001}, it was shown that the inclusive $\alpha$, triton and deuteron spectra have major contributions at forward emission angles. It has been demonstrated that $\alpha$ particles are detected in significantly larger number than the other fragments taken together (deuterons + tritons), which indicates the dominance of incomplete fusion in comparison to elastic breakup at energies near the barrier. 

Due to the difficulty of evaluating the different processes involved in these reactions, a theory able to describe the largest number of accessible states is sought. Over the last few decades, much effort was employed to describe the simpler elastic breakup (EBU) channel, while little emphasis was given to the more complex nonelastic breakup (NEB) contributions. In the NEB emissions, all possibilities for fragment absorption and target final states should be included. 
Theories were developed in the 1980s by several groups in which closed-form expressions were presented for the sum over all possible final states. 
Formalisms for inclusive breakup were developed by: Baur {\it et al.} \cite{baur1980}, as a distorted-wave Born approximation (DWBA) sum-rule; Hussein and McVoy~\cite{hussein1985}, with the extraction of singles emission cross sections in a spectator model with a sum over final states; Udagawa and Tamura (UT), using the prior-form DWBA~\cite{udagawa1981,udagawa1984}; and Ichimura, Austern, and Vincent (IAV), using the post-form DWBA~\cite{ichimura1985,austern1987}. 
An extensive analysis of the equivalence of post-prior theories was reported recently by Lei and Moro for breakup~\cite{lei2015a} and transfer~\cite{lei2015b} reactions.  
These works help to substantiate the success of the post-form DWBA description, which has been corroborated by Carlson {\it et al.}~\cite{carlson2016,carlson2017,hussein2017} and Potel {\it et al.}~\cite{potel2017}.

The calculation of three-body reaction models is by itself a very complicated problem. Generally speaking, in this approach each physical state is composed of a complex superposition of many final reaction states. Four-body like approaches have been investigated lately as extensions of the continuum-discretized coupled-channels (CDCC) method with a low breakup threshold\cite{gallardo2009,matsumoto2010,gallardo2011,descouvemont2018}. CDCC successfully describes the elastic breakup components of reactions with weakly bound three-body projectiles. However, many reactions with loosely bound nuclei are dominated by the incomplete fusion component,i.e, the nonelastic one, which can be calculated using the post-form DWBA. Implementation of a four-body reaction model within the DWBA will require extension of the energy partition between the fragments and of their spin and orbital angular momentum coupling, as well as the consideration of an increased number of possible outgoing channels. This subject is discussed in more detail in Ref. \cite{carlson2017}.

The present paper appraises the post-form IAV DWBA method in calculations of $\alpha$ inclusive cross sections in comparison with available data.
We have implemented the S\~ao Paulo optical potential (SPP) for the different entrance channels. The SPP is composed of a velocity-dependent function multiplied by the folding potential that takes into account the nuclear matter density distribution~\cite{chamon1997,chamon1998, chamon2007}. The diffuseness parameter of the SPP is adjusted here to provide a better description of the data. The absorptive part is assumed to have the same shape as the real potential with a renormalized strength. Standard phenomenological optical models are implemented to calculate the distorted wave functions of the fragments in the final state.  
%
%

This paper is organized as follows: in Sec. \ref{sec:formalism} the IAV DWBA post-form representation is succinctly presented. In section \ref{sec:results}, our calculations are shown for inclusive breakup emissions and compared to the data for several reactions involving $^6$He (Sec. \ref{sec:results6he})  and $^{6,7}$Li (Secs. \ref{sec:results6li}-\ref{sec:results7li}). Finally, Sec. \ref{sec:conclusion} contains the summary and conclusions.

\section{Inclusive Three-Body Breakup Formalism}\label{sec:formalism}

We provide here only the ingredients necessary for the cross section calculations. A more complete description of the inclusive breakup formalism and the underlying assumptions can be found in papers by Ichimura, Udagawa, Austern, Vincent and Kasano  \cite{austern1981,kasano1982,ichimura1985,udagawa1981,ichimura1990}, revisited more recently by Carlson {\it et al.} \cite{carlson2016,carlson2017,hussein2017}, Potel {\it et al.} \cite{potel2015,potel2017} and Lei and Moro \cite{lei2015a,lei2015b}.
Most of the recent works illustrate the success of this approach in calculations of emissions from deuteron breakup.

Consider the reaction $A(a, b)X$ with a two-cluster projectile ($a=b+x$), in which particle $b$ (fragment) is detected and $X$ represents a final state of the fragment $x$ together with the target ($x+A$).  
The inclusive particle emission cross section is generally obtained as a sum of two components
\begin{eqnarray}\label{cs1}
  \frac{d^{2}\sigma ^{\text{}}}{d \Omega_{b} dE_{b}} &=& \frac{d^{2}\sigma ^{\text{EBU}}}{d \Omega _{b}dE_{b}} + \frac{d^{2}\sigma ^{\text{NEB}}}{d \Omega _{b} dE_{b}},
\end{eqnarray}
namely, the elastic (EBU) and nonelastic (NEB) breakup contributions. 
The contribution to the elastic mode is written in terms of its T-matrix element as
\begin{multline}
  \frac{d^{2}\sigma ^{\text{EBU}}}{d \Omega _{b} dE_{b}} = 
  -\frac{2}{\hbar v_{a}} \rho_{b}\left(E_{b}\right)
  \left|T\left(\mathbf{k}_{b}, \mathbf{k}_x ; \mathbf{k}_{a}\right)\right|^{2}\\ \times \delta\left(E_{a}+\varepsilon_{b}-E_{b}-E_{x}\right),
\end{multline}
  with the final momentum density $$ \rho_b(E_{b})=\frac{m_{b} k_{b}}{8\pi^3 \hbar^2}.$$
  The T-matrix element in the post-form is given by
  \begin{multline}
  T\left(\mathbf{k}_{b}, \mathbf{k}_x ; \mathbf{k}_{a}\right)=
    \\ =\langle \tilde {\chi _{b}}^{(-)}(\mathbf{k}_b,\mathbf{r}_b) \tilde {\chi _x}^{(-)}(\mathbf{k}_x,\mathbf{r}_x) |V_{b x}(\mathbf{r}) |\chi _{a}^{(+)} (\mathbf{k}_a,\mathbf{R}) \phi_a (\mathbf{r})\rangle,
  \end{multline}
where $(+)$ ($(-)$) represents the incoming (outgoing) scattering wave function. We approximate the ground state $\phi_a$ of the projectile together with the two-body interaction $V_{b x}$ by taking the zero-range approximation (ZR) to write,
\begin{eqnarray}
 D_{0} = \int d\mathbf{r}\,  V_{b x}( \mathbf{r})\, \phi_a (\mathbf{r}).
\end{eqnarray}
The strength $D_{0}$ will be adjusted by comparison with the experimental data. 
In the contact interaction approach, the T-matrix becomes
\begin{multline}
 T\left(\mathbf{k}_{b}, \mathbf{k}_x ; \mathbf{k}_{a}\right) =
 \\
=
 D_{0}\langle \tilde {\chi _{b}}^{(-)}(\mathbf{k}_b,\mathbf{r}_b) \tilde {\chi} _{x}^{(-)} (\mathbf{k}_x,\mathbf{r}_x)
 \Lambda(r_x) | {\chi} _{a}^{(+)} (\mathbf{k}_a,\mathbf{R} )  \rangle ,
\end{multline}
where $\Lambda(r_x) $ accounts for the effects of finite-range corrections (FR)  {} {(discussed in details at Cap.~6 of Ref.~\cite{Satchler1983}). 
} 
 
The nonelastic contribution is obtained from the imaginary part of the optical potential between the absorbed fragment and the target as
\begin{multline}
  \frac{d^{2}\sigma ^{\text{NEB}}}{d \Omega _{b} dE_{b}}  =  
  -\frac{2}{\hbar v_{a}} \rho_{b}\left(E_{b}\right) 
  \\
  \times
  \langle \Psi _x (\mathbf{k}_b , \mathbf{r}_x ; \mathbf{k}_a) |W_{x}  (\mathbf{r}_x)| \Psi _x (\mathbf{k}_b , \mathbf{r}_x ; \mathbf{k}_a) \rangle ,  
  \end{multline}
 where,\bibliographystyle{ametsoc}
\begin{equation}  
  | \Psi _x \rangle = D_{0 }\left ( \tilde {\chi _{b}}^{(-)} (\mathbf{k}_b,\mathbf{r}_b)
  G^{(+)} _x(\mathbf{r}_x, \mathbf{r}_{x}^\prime) \Lambda(r_x) |\chi_{a}^{(+)}(\mathbf{k}_a, \mathbf{R})   \right \rangle 
\end{equation}
is the effective wave function, in the zero-range approximation, for the propagation of the remaining fragment by the optical model Green`s function $G^{(+)}$. 

The finite (FR) and zero range (ZR) approximations have been extensively compared in the context of deuteron breakup in Potel  {\it et al.}~\cite{potel2017}. The results obtained with both approaches were found to be in good agreement. Lei and Moro \cite{lei2015a}, have also provided a comparison between FR and ZR calculations. They performed calculations of the deuteron-induced $^{58}$Ni($d$,$p$X) cross section at 80 MeV (see Figure 3 of \cite{lei2015a}) and of the alpha emission cross section from $^6$Li+$^{209}$Bi at the incident energies of 24 MeV and 38 MeV (see Figure 7 of \cite{lei2015a}) and compared the NEB component of the two approaches. One can see in Figure 7 that the ZR DWBA calculations underestimate the FR results by about $10\% - 20\%$. However the overall behaviour of the distributions is similar. The difference between the two approximations thus does not constitute an obstacle to employing the ZR method, since the contact term is parameterized by a strength $D_0$ which can always be fitted to provide the proper cross section.
 
We perform numerical calculations by expanding the wave functions and T matrix elements in partial waves as in Ref.~\cite{carlson2016}. 
The nucleon distorted waves were obtained with the Koning-Delaroche optical potentials \cite{koning2003}, while for the deuteron(dineutron)-target we used a deuteron optical potential \cite {haixia2006}. For the projectile-target interaction, we employ the Sao Paulo optical potential (SPP)~\cite{chamon1997,chamon1998, chamon2007}.

The wave function for the entrance channels were obtained using the SPP. 
Although the SPP was developed principally for heavy-ion interactions, it has performed very well for light-nucleus-induced reactions,as shown in a number of studies~\cite{zagatto2017,gasques2018,alvarez2018}. 
We performed a best fit of a Wood-Saxon function to the real part of the SPP and used the resulting strength and geometrical parameters in our calculations. Both the imaginary and the real potential have the same shape with the exception of their depths, where ${W_0=0.78~V_0}$ is used, following the usual SPP systematic. For the cases studied in this paper, the SPP produces a strongly diffusive potential ${a_0^{\textrm{SPP}} \approx 1 \text{~fm}}$ resulting in a very absorptive surface for the imaginary part. We follow Ref.~\cite{chimanski2021} and use a smaller diffuseness parameter, ${a_0=0.65 \text{~fm}}$, for both the real and imaginary components, a value in accordance with the phenomenology of light-nucleus optical potentials. The parameters employed in our calculations are energy independent and are given in Table~\ref{tab:parametersSPP}.

\begin{table}[ht]
\centering 
\caption{Optical potential parameters for the different entrance channels studied. The parameters are obtained by fitting a Wood-Saxon function to the SPP (see text for details). We restrain the imaginary and real parts by taking ${W_0=0.78~V_0}$.  The geometry for the imaginary part is taken to be the same os the real one $r_{i_0}=r_0$.}
\begin{tabular}[t]{c||c|c|c|c|c}
\hline 
Reaction  & $E_a$(MeV)& $V_0$(MeV) & $W_0$(MeV) & $a_0^{\textrm{}}$(fm) & $r_0$ (fm) \\
\hline\hline
$^6$He+$^{64}$Zn & 14.85 &  -272.0 &  -212.1  &  0.65 &  1.108\\
\hline
$^6$He+$^{120}$Sn & 18.0 & -275.7 &  -215.0   & 0.65  & 1.146\\
\hline
$^6$He+$^{209}$Bi & 18.0 & -274.4  &  -214.0  &  0.65  & 1.177\\
\hline\hline  
$^6$Li+$^{58}$Ni & 18.0 & -273.0 & -212.9  & 0.65 & 1.104\\
\hline
$^6$Li+$^{90}$Zr & 18.0 &  -278.7   & -217.4  &  0.65 &  1.126\\
\hline
$^6$Li+$^{118}$Sn & 24.0 & -280.5  & -218.8 &    0.65  &  1.143\\
\hline
$^6$Li+$^{208}$Pb & 30.0 & -280.6 &  -218.9 & 0.65  & 1.175\\
\hline\hline
$^7$Li+$^{56}$Fe & 68.0 & -309.8 & -241.7 & 0.65 & 1.105\\
\hline
$^7$Li+$^{58}$Ni & 17.5 & -313.6  &  -244.6  &  0.65 &   1.106\\
\hline
\end{tabular}
\label{tab:parametersSPP}
\end{table}%
For the light fragments, deuteron, dineutron, triton and $\alpha$, we turned to standard phenomenological optical potentials from the literature. These interactions are composed of Woods-Saxon like functions for the real and imaginary parts, including both volume and surface terms.  

We take the deuteron optical potential parameters from the work of Han {\it et al.} Ref.~\cite{han2006}, a potential developed for targets in the mass range ${12\leqslant A \leqslant 209}$ with incident energies from threshold up to 200 MeV. This global optical potential is recommended for use in deuteron induced reactions in the Reference Input Parameter Library (RIPL), Sec. F of Ref.~\cite{capote2009}. According to the same reference, the most appropriate potentials for our particular applications would be those we have used - the global potential of Becchetti and Greenlees for triton and of Avrigeanu {\it et al.} for $\alpha$ particles.

For convenience we list the optical potentials employed here in Table~\ref{tab:opticalpot}.  
We model the dineutron optical potential (in the exit channel of the breakup of $^6$He) as a deuteron-like interaction with zero charge and spin. We will come back to this approximation below.

\begin{table}[ht]
\centering 
\caption{Optical potentials employed for fragments in the reaction exit channels. The dineutron is taken as chargeless deuteron with zero spin.}
\begin{tabular}[t]{c|c}
\hline 
Particle  & Optical potential  \\
\hline\hline
 deuteron & Han-Shi-Shen~\cite{han2006} \\
\hline
 dineutron & Han-Shi-Shen (adapted)~\cite{han2006}\\
\hline
triton & Beccheti-Greenless~\cite{becchetti1969} \\
\hline 
 $\alpha$ & Avrigeanu et al.~\cite{avrigeanu1994}\\
\hline
\end{tabular}
\label{tab:opticalpot}
\end{table}%

%

\section{Results and Discussion}\label{sec:results}
  In this section we present inclusive breakup cross section calculations for different systems. We first investigate reactions induced by $^6$He on $^{64}$Zn, $^{120}$Sn and $^{209}$Bi targets. The lithium projectiles $^6$Li and $^7$Li incident on mid-weight and heavy targets $^{58}$Ni,$^{90}$Zr, $^{118}$Sn, $^{208}$Pb and $^{56}$Fe,$^{58}$Ni, are analyzed subsequently. For convenience, we have compiled in Table~\ref{tab:D0s}, the contact strength  $D_0$ and the binding energies employed for the reactions studied.

\subsection{$\alpha$ emission from the breakup of the halo nucleus \textsuperscript{6}He}
\label{sec:results6he}
We have reduced the problem of $^6$He from a three- to a two-body one by describing the projectile structure as a composition of $\alpha$ plus a dineutron particle.  

The possibility of considering a dineutron approach was first proposed in 1972 by Migdal \cite{migdal}. Even though the force between the two neutrons is not strong enough to create a bound system, the existence of an extremely short-lived resonance state produced during the nuclear reaction was proposed, in which the two neutrons form a composite system with a core nucleus. This is the formation of a three-body bound nucleus in which any two-body subsystem is unbound, namely, a Borromean system. Considerable experimental effort has been made in order to investigate such two-neutron halo nuclei~\cite{tanihata1985,balamuth1994,aksyutina2009,brodeur2012}. Theoretical work also follows these lines \cite{gallardo2009,baye2009,gallardo2011,matsumoto2010,frederico2012} in exploring the existence of the dineutron. The first observation of dineutron emission was found in the decay of $^{16}$Be  \cite{Spyrou2012} where the three decay products ($^{14}$Be+$n$+$n$) were measured in a single-proton knockout reaction. Also measured was the averaged correlation angle between the two neutrons $\langle \theta_{nn}\rangle$. In two very recent works, from Kubota {\it et al.}~\cite{kubota2020} and from Sun {\it et al.}~\cite{sun2021}, the formation of a dineutron in the $^{11}$Li nucleus ($^{9}$Li+$n$+$n$) was observed. In Ref.~\cite{kubota2020}, for the first time, $\langle \theta_{nn}\rangle$ was measured as a function of the intrinsic neutron momentum, with the observed distribution equivalent to a localization of the dineutron at $\approx$3.6 fm from the core. In Ref. ~\cite{sun2021}, the Coulomb and nuclear breakup of $^6$He on Pb and C targets at 70 MeV/nucleon was studied. The authors found an angle of $\langle \theta_{nn}\rangle \approx 56 $ degrees. This angle, similar to those found in Refs.~\cite{Spyrou2012,kubota2020}, is significantly smaller than the uncorrelated one, $\langle \theta_{nn}\rangle = 90$ degrees. In summary, the measurements show a spatial correlation of the two neutrons in the halo. This motivated the dineutron approximation in the reaction calculations for  $^6$He in this work.

The typical experimental two-neutron separation energy ${S_{2n}=0.975 \text{~MeV}}$ for $^6$He describes a three-body picture fairly well but leads to a spatially expanded wave function in a two-body model. Therefore, we have used the modified binding energy of ${\varepsilon_b=-1.6  \text{~MeV}}$ suggested in Ref.~\cite{moro2007} and compared cross section calculations using both values. The binding energy is the energy required to break  $^6$He into an $\alpha$ and a dineutron (${\varepsilon_b=-S_\alpha}$).

\begin{figure}[h]
\centering
\includegraphics[width=0.92\linewidth]{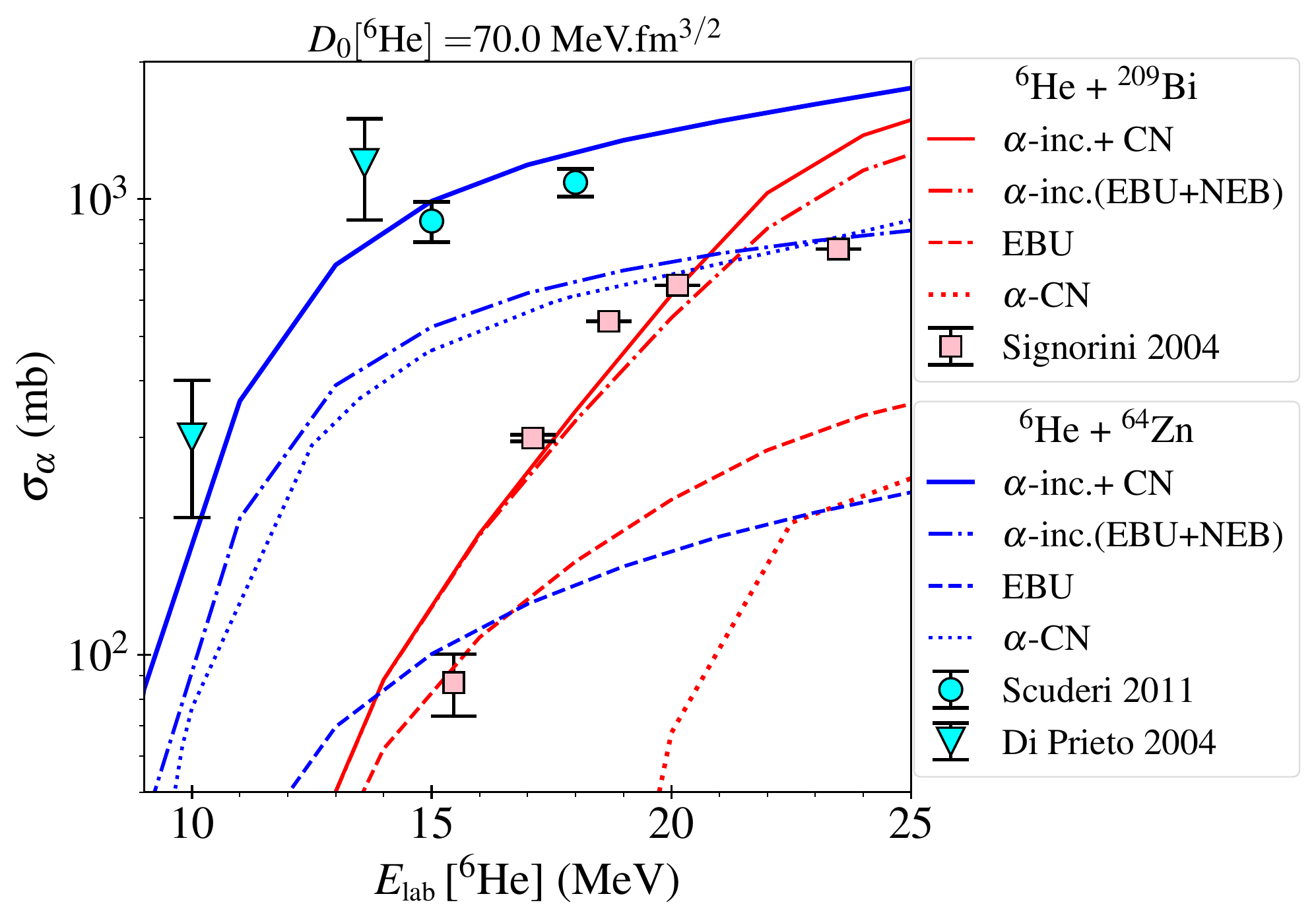}
\vspace{-0.5cm}
\caption{Comparison of inclusive $\alpha$  cross sections for $^6$He incident on $^{209}$Bi  and $^{64}$Zn to the data taken from~\cite{signorini2004} (squares),~\cite{scuderi2011} (circles) and~\cite{dipietro2004} (triangles). Compound nucleus cross sections (dotted lines) are computed using the EMPIRE code~\cite{herman2007}. The binding energy of the projectile is taken as ${\varepsilon_b=- 1.6 \text{~MeV}}$  within the two-body model. See the discussion in the text.}
\label{fig:6He}
\end{figure}

\begin{figure}[h] 
\includegraphics[width=0.85\linewidth]{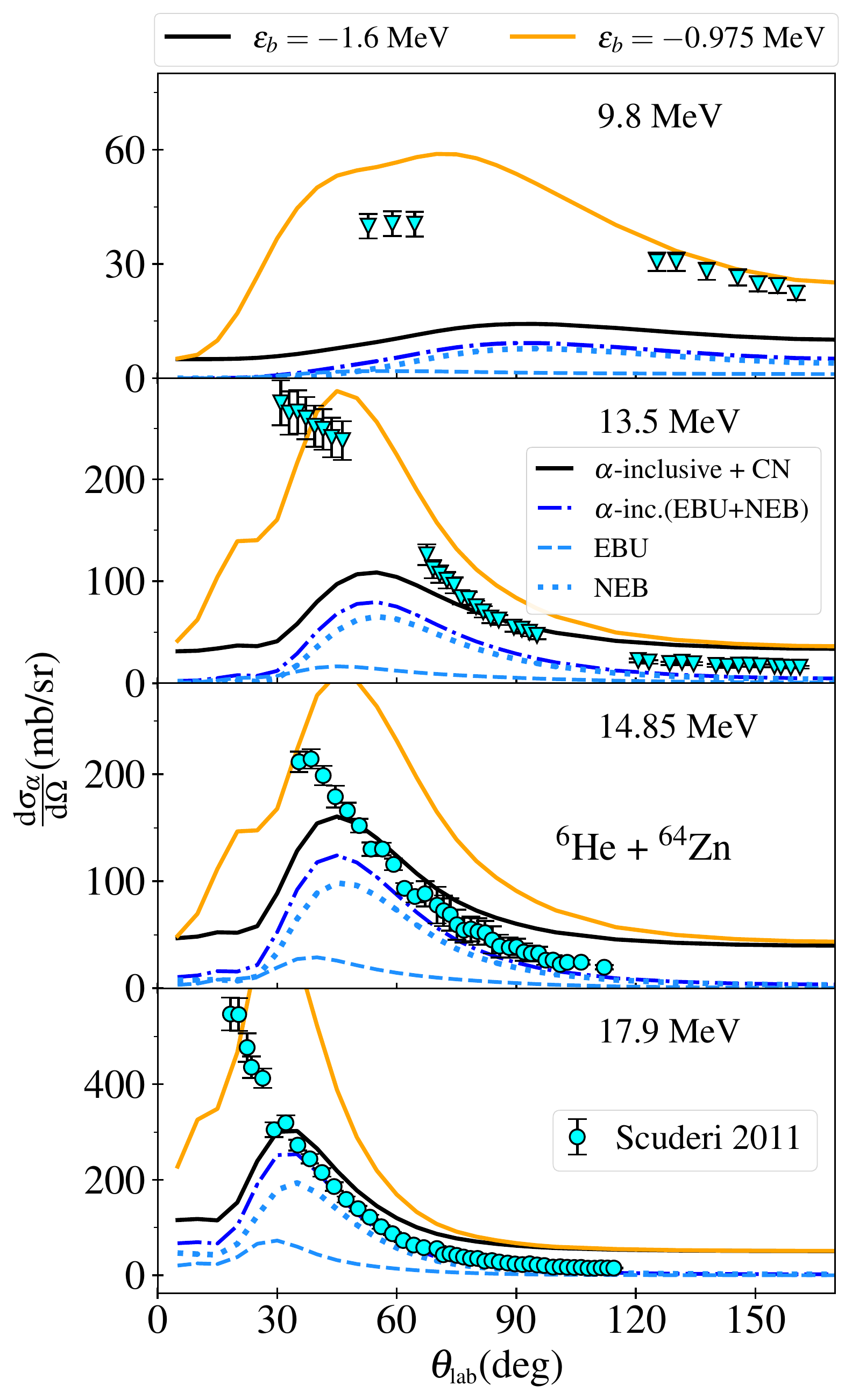}
\vspace{-0.5cm}
\caption{$\alpha$ particle angular distribution from the breakup of $^6$He on $^{64}$Zn at the laboratory energies indicated in the panels. Experimental data are taken from Refs.~\cite{dipietro2004} (circles) and~\cite{scuderi2011} (triangles). }
\label{fig:6He-64Zn}
\end{figure}   

We show in Fig.~\ref{fig:6He} the integrated $\alpha$ particle cross sections for two different targets, $^{64}$Zn and $^{209}$Bi.
 The calculations are compared to experimental data taken from Refs.~\cite{signorini2001} (squares),~\cite{scuderi2011} (circles)
  and~\cite{dipietro2004} (triangles).
The dotted lines represent the contribution from the compound nucleus (CN), calculated with the EMPIRE code~\cite{herman2007}. The CN decay is described using the Hauser-Feshbach model with optical model transmission coefficients, assuming isotropic emission in the center-of-mass  frame and including the possibility of multi-particle emission. We refer to Ref.~\cite{herman2007} for a detailed description of the formulation of the $\gamma$-ray cascade, strength functions and level densities employed. The CN emission cross sections from EMPIRE are added to the inclusive breakup emission cross sections to obtain the total $\alpha$ emission cross section. The inclusive emission cross sections are shown as dashed lines and the sum of the two (CN + INC) is represented by the solid lines.

We have adjusted the value of the zero-range constant ${D_0 =70 \textrm{~MeV.fm}^{3/2}}$ to best reproduce, simultaneously for both targets, the values of the experimental cross section data.
 In the calculations used to determine the $D_0$ value, we have taken the binding energy as ${\varepsilon_b=-1.6 \text{~MeV}}$, so as to reproduce the dineutron model of the {$\alpha$+dineutron} in Ref.~\cite{moro2007}.

In the case of $^6$He+$^{209}$Bi, our calculation provides a reasonable description of the trend shown by the experimental data. A more detailed calculation would involve other reaction components such as transfer, which should also play a role here, as observed by the authors of Ref.~\cite{signorini2004}.

Angular distributions of inclusive $\alpha$ particle emission from $^6$He fragmentation on $^{64}$Zn are shown in Fig.~\ref{fig:6He-64Zn} for several incident energies. The calculations are compared to data from Ref.~\cite{dipietro2004} in the two top panels and to data from~\cite{scuderi2011} in the two bottom panels. 
Here we have studied the effects of the experimental and the modified effective binding energies of $^6$He. 
As mentioned earlier, in Ref.~\cite{moro2007}  the {$\alpha$+dineutron} binding energy must be modified to ${\varepsilon_b= -1.6 \text{~MeV}}$ in order to reproduce the $^6$He single-particle density. The modified binding energy furnishes a good description of the elastic scattering data of $^6$He in the reduced two-body model. We also show cross sections calculated using the experimental two-neutron separation energy of ${\varepsilon_b=-S_{2n}=-0.975 \text{~MeV}}$.
When $S_{2n}$ is taken as the measured binding energy in the {dineutron+$\alpha$} model, the angular distribution is overestimated.


In Fig.~\ref{fig:6He-64Zn}, we show that we obtain a rather poor description of the alpha emission due to breakup at 9.8 MeV, when using ${\varepsilon_b=-1.6 \text{~MeV}}$, a region where one might expect other processes to dominate the emission.
In general, we have found a better agreement between theory and experiment at large angles for incident energies increasing above ${13.5  \text{~MeV}}$.   
The data are extracted from the two references~\cite{dipietro2004} and \cite{scuderi2011}. A similar method for obtaining the reduced data is used in both experiments. Based on a three-Gaussian fit, the elastic and inelastic scattering peak are subtracted from a broad $\alpha$-emission bump.
In the data reduction, forward angles (in the region with $\theta<30^\circ$) were excluded, as the elastic scattering contribution was too strong to be subtracted, which partially justifies the discrepancy between the data and our calculation. 

We found our calculations to perform better at higher energies. The discrepancies at small angles can be associated with the ambiguities and challenges faced in the data extraction. From the theoretical point of view, the forward scattering angles also provide a challenge to our model, since we do not take into account all possible components involved in the reaction. Here we note that more experimental data is needed to clarify the reaction mechanisms responsible for emission at small angles.

\begin{figure}[!htb] 
\includegraphics[width=0.85\linewidth]{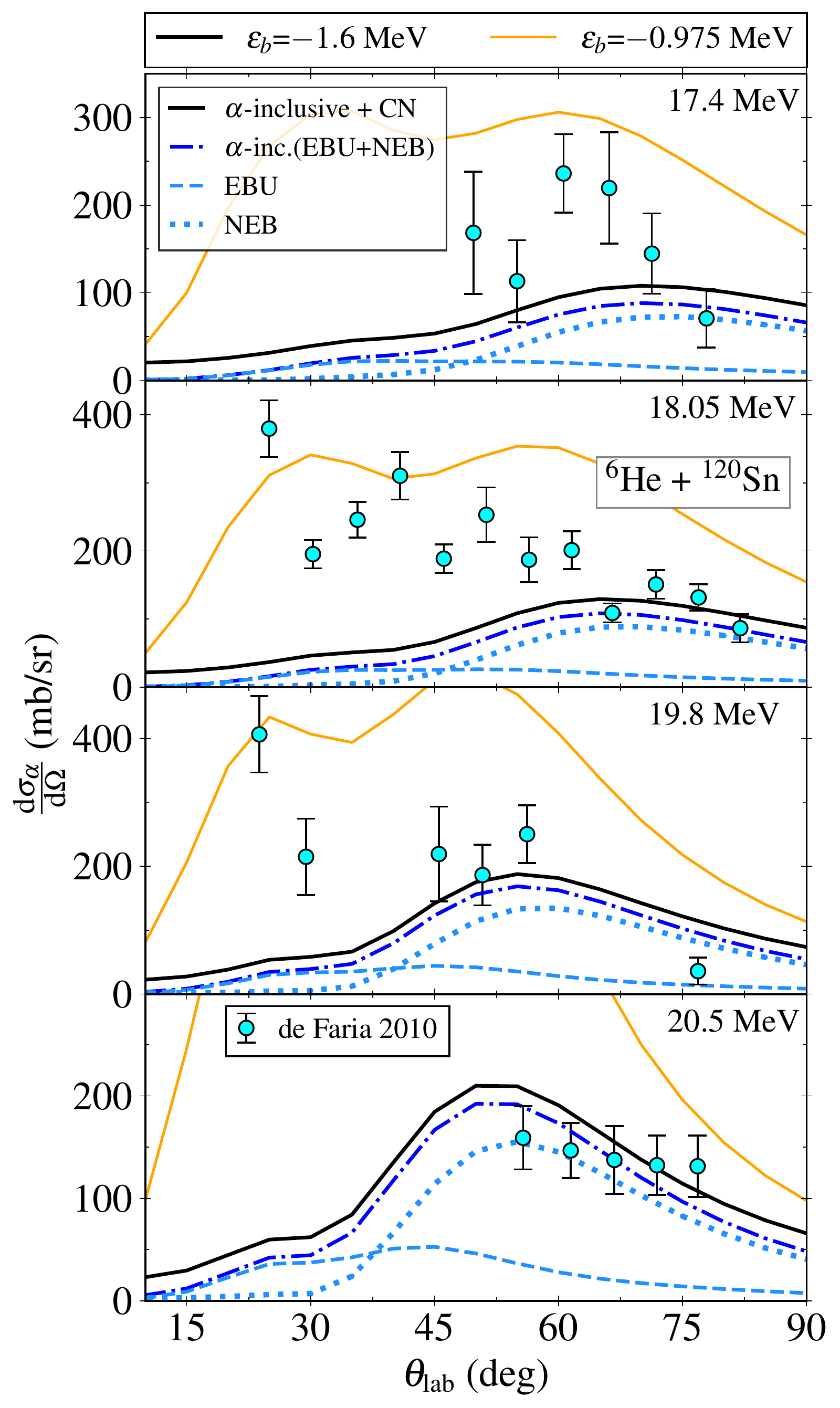}
\caption{Inclusive angular distribution for $\alpha$ particle emission in the reaction $^{120}$Sn($^6$He,$\alpha$)X at the energies indicated in the frames. The angular distributions obtained using the experimental two-neutron separation energy in the $\alpha$+dineutron model are also shown (orange lines). The experimental data were taken from Ref.~\cite{defaria2010}.
}
\label{fig:defaria2010}
\end{figure}

Inclusive angular distributions for $\alpha$ emissions have been measured in the reaction $^{120}$Sn($^6$He,$\alpha$)X, and are compared to our calculations in Fig.~\ref{fig:defaria2010}. Our predictions tend to underestimate the experimental data for this case. We point out that the results obtained here do not differ significantly from the DWBA calculations presented in the work of P. N. de Faria and collaborators (Ref.~\cite{defaria2010}), from which the experimental data of Fig.~\ref{fig:defaria2010} were extracted.

\begin{figure}[!]
\centering
\includegraphics[width=0.97\linewidth] {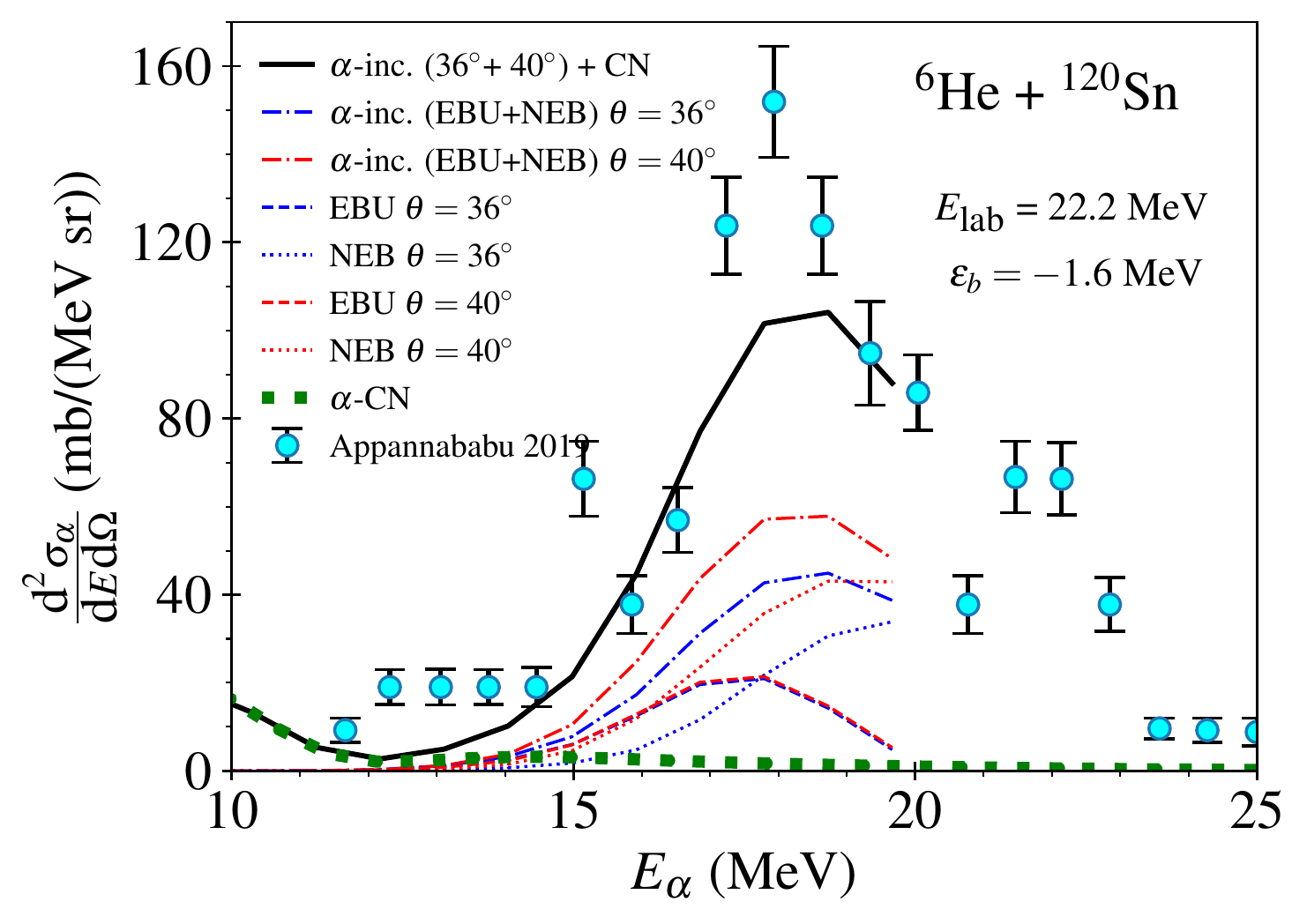}
\vspace{-0.5cm}
\caption{Double differential $\alpha$ emission cross section from the reaction $^{120}$Sn($^6$He,$\alpha$)X at a laboratory energy of ${22.2 \text{~MeV}}$ for a sum of two different angles, $\theta_{\textrm{lab}}=36^\circ$ and $\theta_{\textrm{lab}}=40^\circ$. The experimental data are taken from Ref.~\cite{appannababu2019}}
\label{fig:6He-120Sn-spectra}
\end{figure}

%
A double differential cross section for the same reaction but at 22.2 MeV is shown in Fig.~\ref{fig:6He-120Sn-spectra}. In this case, the cross section is obtained by summing the measured differential spectra at two scattered angles, $\theta_{\textrm{lab}}=36^\circ$ and $\theta_{\textrm{lab}}=40^\circ$. The distributions increase as the energy rises above the Coulomb barrier of about {$V_B\approx 13.3 \text{ MeV}$} \cite{freitas2016,pires2018}. The differential spectra obtained using the experimental two-nucleon separation energy (not shown here) overestimate the data by more than a factor of two.
We also note that our model does not include negative energies of the absorbed fragment (stripping), which limits the calculated curves to energies below $E_a+\varepsilon_b$.

\begin{table}[!htb]
\centering 
\caption{Properties of the projectiles studied: contact strength parameters $D_0$ and binding energies $\varepsilon_b$. We have used two different binding energies in the dineutron model of $^6$He, one being the effective binding energy of the two-body model ($\varepsilon_b=-1.6$ MeV~\cite{moro2007}), and the other the experimental two-neutron separation energy ($\varepsilon_b=-S_{2n}=-0.975$ MeV). The latter does not reproduce the two-body dineutron nucleon density well and was implemented only to show the large cross sections it furnishes when used in the model.}
\begin{tabular}[t]{c|c|c}
\hline 
Nucleus \ \  & $D_{0}$ (MeV.fm$^{3/2}$) &   $\varepsilon_b$ (MeV)\\
\hline\hline
$^6$He  & 70.0 & -0.975 and -1.6\\
\hline 
$^6$Li  & 45.0 & -1.474\\
\hline 
$^7$Li  & 42.0 & -2.468\\
\hline
\end{tabular}
\label{tab:D0s}
\end{table}%


\begin{figure*}[!] 
\centering
\includegraphics[width=0.7\linewidth]{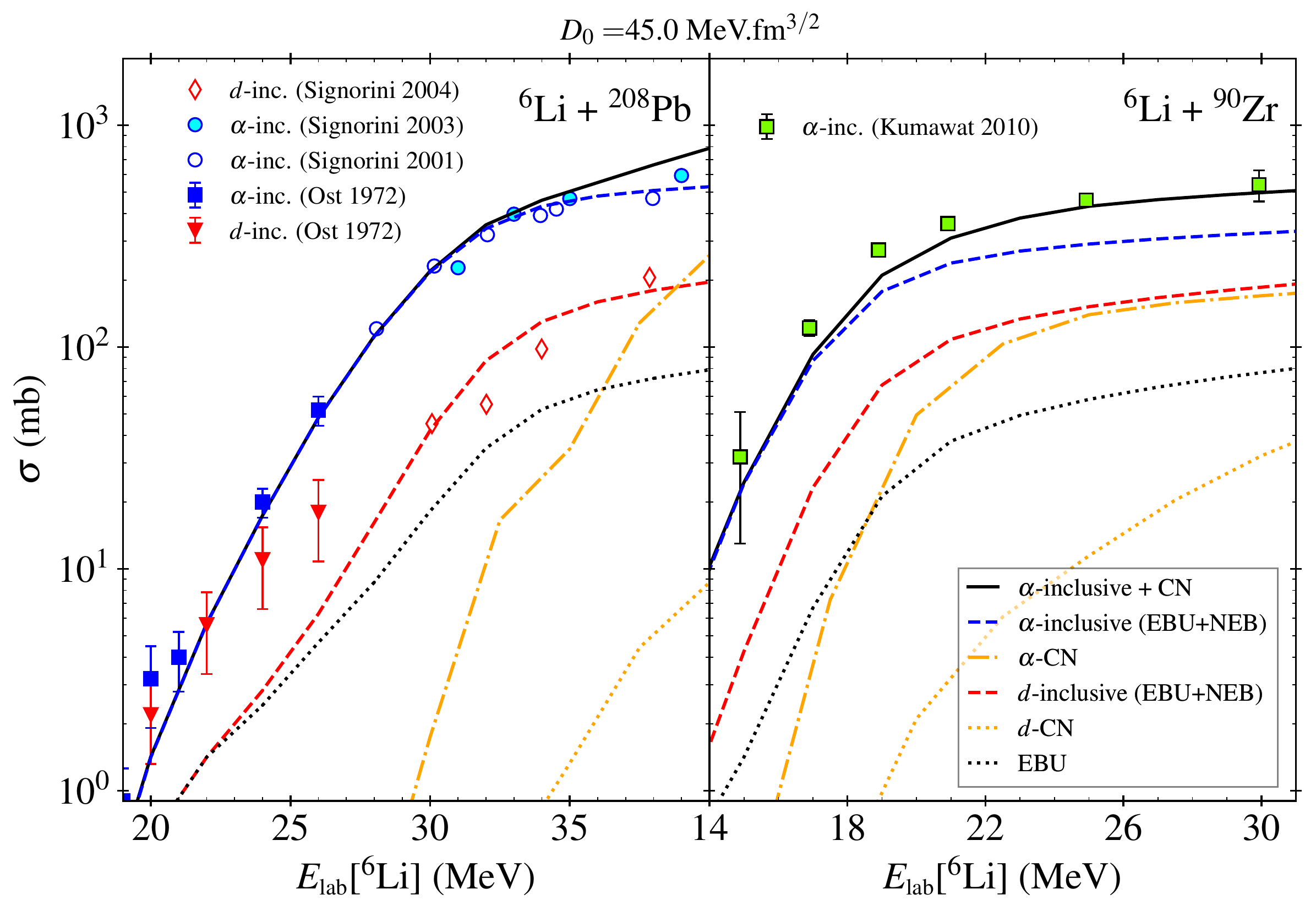}
\vspace{-0.5cm}
\caption{Inclusive cross sections for $\alpha$ and deuteron emission.
The experimental data in the panels are taken from Refs.~\cite{ost1972,signorini2001,signorini2003,signorini2004} for $^6$Li+$^{208}$Pb, and~\cite{kumawat2010} for $^6$Li+$^{90}$Zr. Compound nuclear reactions are computed using EMPIRE-III. See text for details.}
\label{fig:6Li}
\end{figure*}

\subsection{$\alpha$ emission from the breakup of \textsuperscript{6}Li }
\label{sec:results6li}

Reactions induced by $^{6}$Li tend to present large $\alpha$ and deuteron emission cross sections. This is expected, as the low $\alpha$+deuteron separation energy ($\varepsilon_b=-1.474$ MeV) facilitates the breakup process. In Fig.~\ref{fig:6Li}, we present integrated alpha and deuteron cross sections for $^{6}$Li induced reactions on $^{208}$Pb and $^{90}$Zr. The compound nucleus contributions to the emissions are also shown.   
The emission cross sections for the $^{208}$Pb target (left panel) are compared to a set of experimental data from different publications. The triangles and squares are from Ref.~\cite{ost1972}, open and filled circles from ~\cite{signorini2001} and~\cite{signorini2003}, respectively, while diamonds identify inclusive deuteron cross sections taken from~\cite{signorini2004}. In the right panel, we compare the calculations for $^{90}$Zr to the experimental $\alpha$ emission data from Ref.~\cite{kumawat2010}. The elastic breakup emission, shown as the dotted black lines in both panels, accounts for only a small part of the cross section. Most of the emitted particles come from the nonelastic component. The zero-range constant $D_0[^6\textrm{Li}]=45 \textrm{ MeV.fm}^{3/2}$ adjusts the solid curves to the experimental data simultaneously for both targets.

\begin{figure}[!] 
\centering
\includegraphics[width=0.85\linewidth] {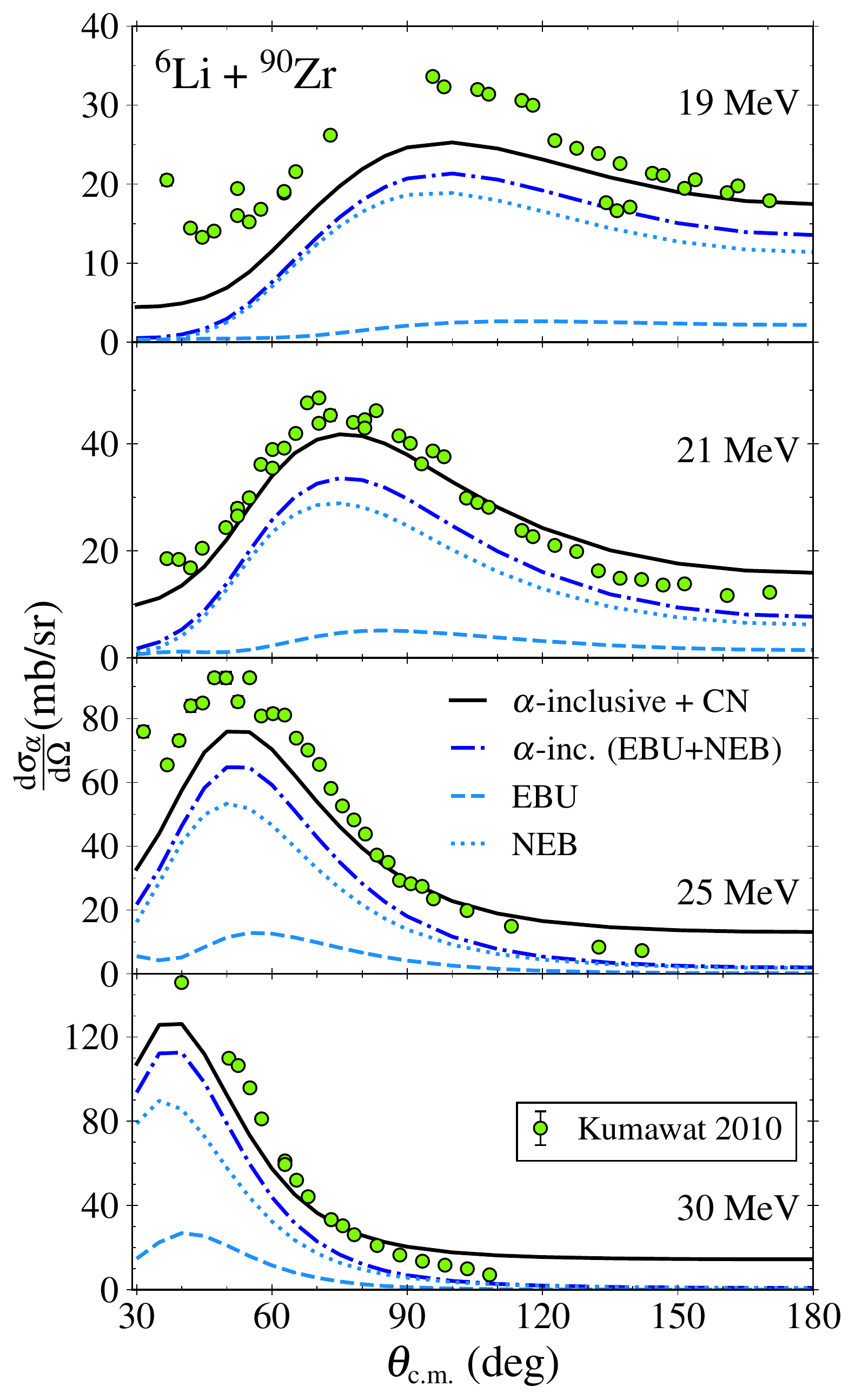}
\vspace{-0.5cm}
\caption{Angular distribution of $\alpha$ particles from the reaction  $^{90}$Zr($^6$Li,$\alpha$)X.
The experimental data are taken from Ref.~\cite{kumawat2010}.}
\label{fig:6Li-90Zr}
\end{figure}

\begin{figure}[!htb]
\centering
\includegraphics[width=0.9\linewidth] {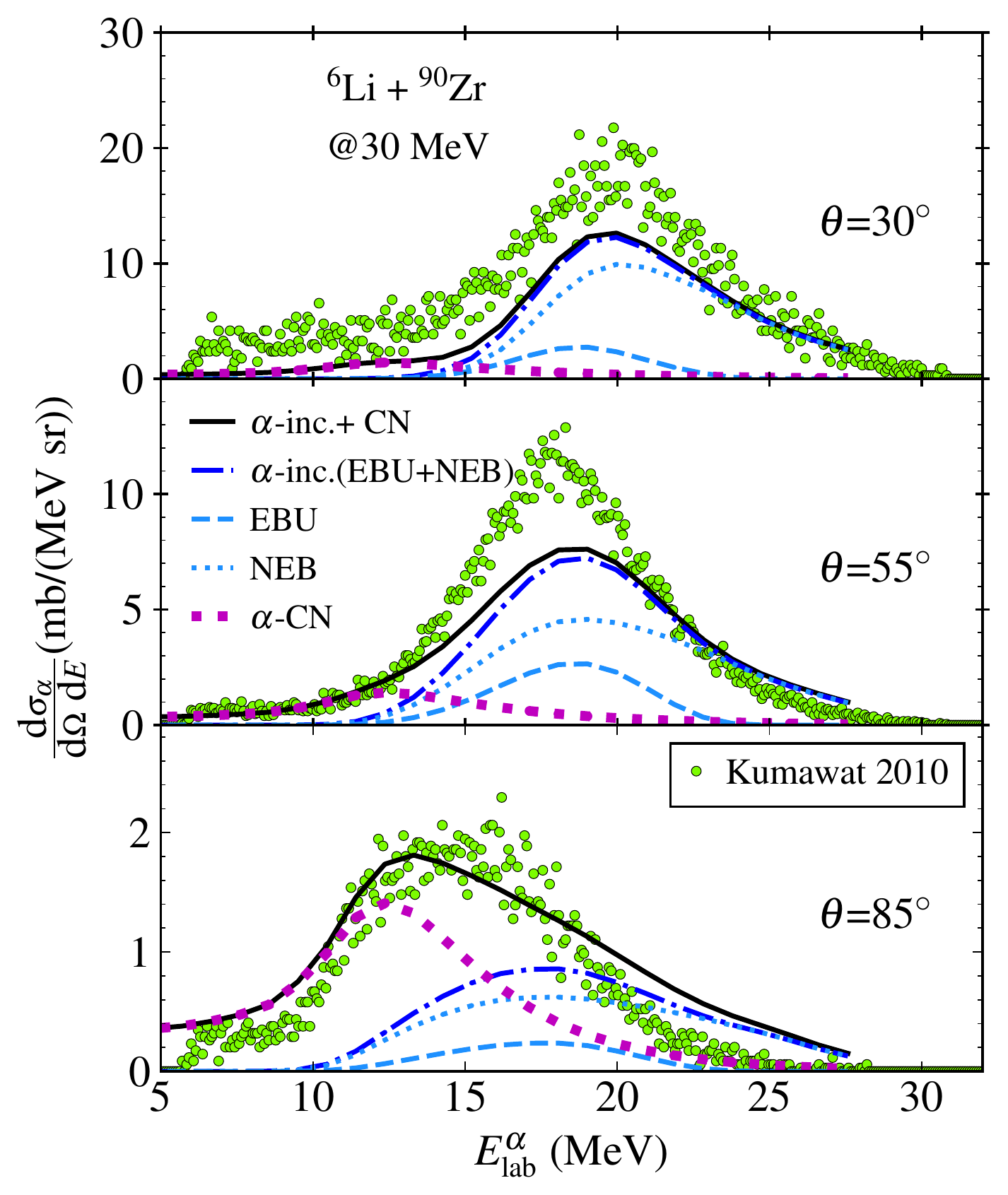} 
\vspace{-0.5cm}
 \caption{Differential energy spectrum of $\alpha$ particles from the breakup of  $^6$Li on $^{90}$Zr at three different angles indicated in the frames. The dashed lines represent the elastic breakup. The experimental data are taken from Ref.~\cite{kumawat2010}.}
\label{fig:6Li-90Zr-spectra}
 \end{figure}

We now turn to the alpha angular and double differential distributions presented in Fig.~\ref{fig:6Li-90Zr} and Fig.~\ref{fig:6Li-90Zr-spectra}, respectively. Our calculations are compared to the experimental data taken from Ref.~\cite{kumawat2010}. Angular distributions (Fig.~\ref{fig:6Li-90Zr}) are compared for several different incident energies. We find an overall good agreement in all cases when the contribution from compound nuclei emission, assumed to be isotropic, is taken into account.
In Figure~\ref{fig:6Li-90Zr-spectra}, we show the differential alpha emission spectrum for three different detection angles. The mean value of the calculated $\alpha$ distributions is well reproduced for $\theta = 30^\circ$ and $\theta = 55^\circ$. The emission at the more backward angle of $\theta = 85^\circ$ is dominated by compound nucleus evaporation, as shown by the pink dotted line.


\begin{figure}[!htb]
\centering
\includegraphics[width=0.9\linewidth] {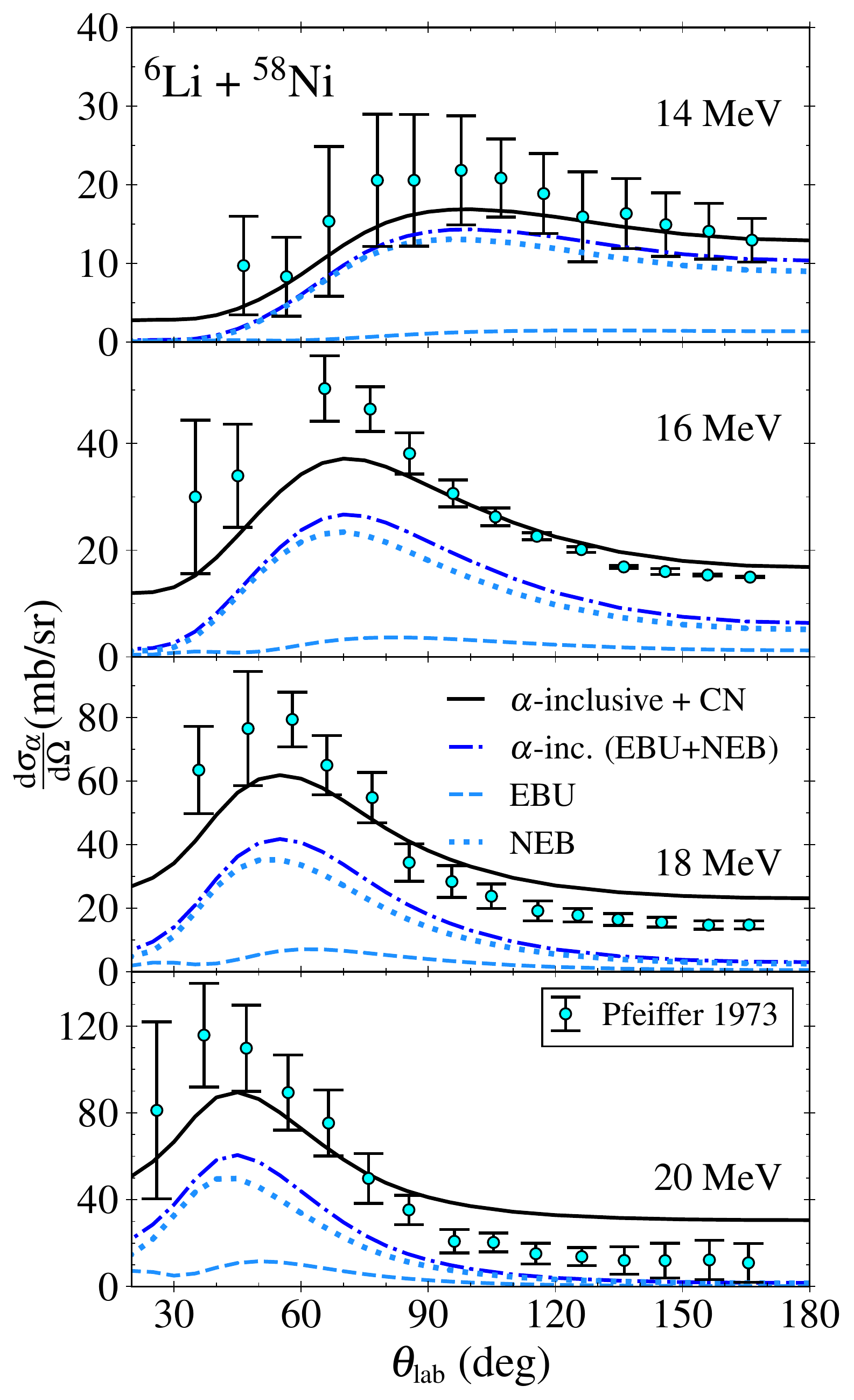} 
\vspace{-0.5cm}
 \caption{Inclusive $\alpha$ angular distribution in the $^6$Li+$^{58}$Ni reaction. The contribution of compound nucleus emission was assumed to be isotropic. The experimental data are taken from Ref.~\cite{pfeiffer1973}.}
\label{fig:6Li-58Ni}
 \end{figure}

\begin{figure}[!htb]
\centering
\includegraphics[width=0.9\linewidth] {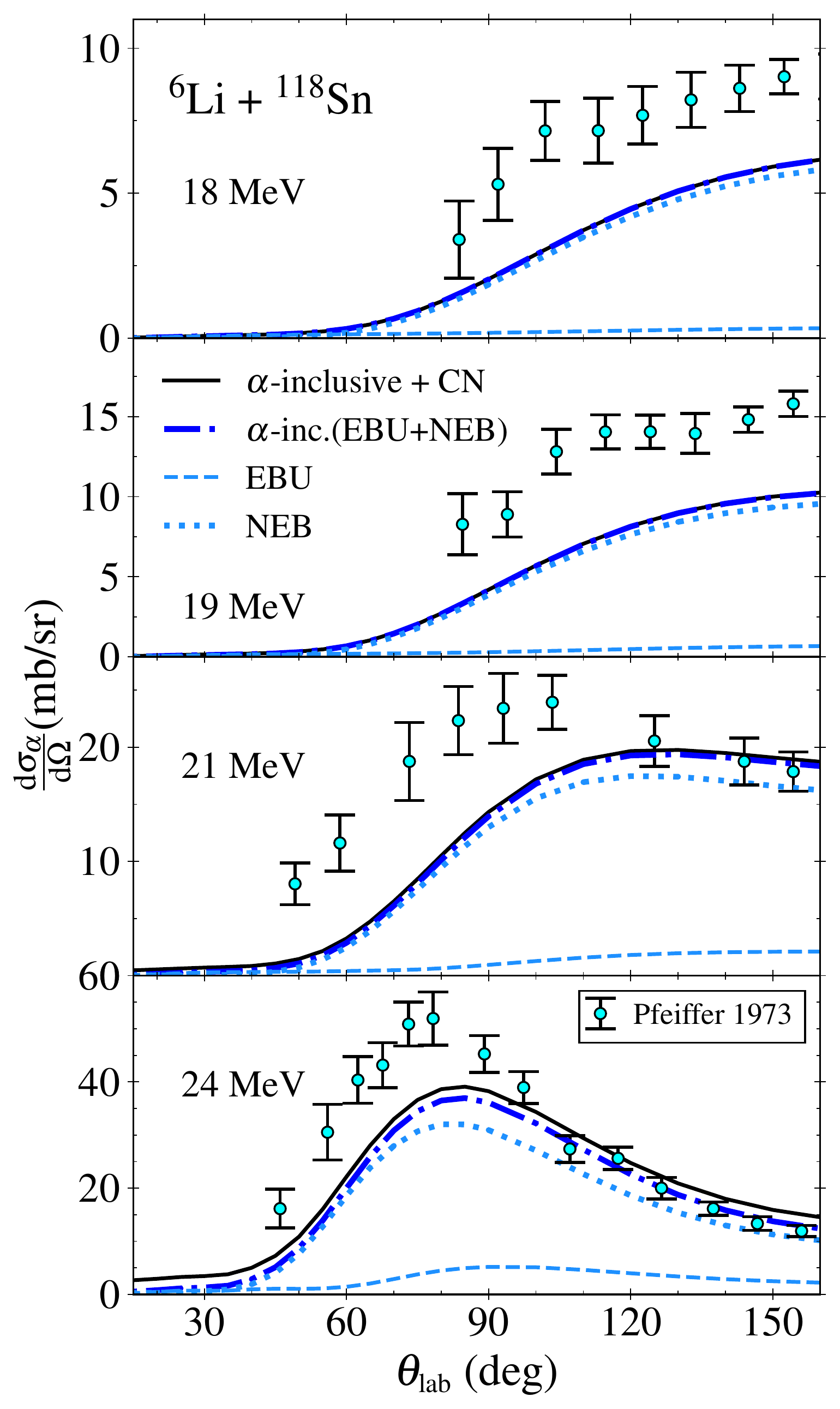} 
\vspace{-0.5cm}
 \caption{Inclusive $\alpha$ angular distribution in the $^6$Li+$^{118}$Sn reaction. The small contribution of the compound nucleus was assumed to be isotropic. The experimental data are taken from Ref.~\cite{pfeiffer1973}.}
\label{fig:6Li-118Sn}
 \end{figure}

 We have also analyzed the $\alpha$ angular distributions from breakup of $^6$Li colliding with $^{58}$Ni and $^{118}$Sn. The results are depicted in Fig.~\ref{fig:6Li-58Ni} and Fig.~\ref{fig:6Li-118Sn}, respectively. The experimental data are from the work of Pfeiffer {\it et al.}, Ref.~\cite{pfeiffer1973}. Our results follow the shape of the experimental distributions very well for both targets. The compound nucleus emission is shown to be important in the $^6$Li + $^{58}$Ni reaction. In the case of the $^{118}$Sn target, due to its larger charge, the contribution of evaporation to the $\alpha$ emission cross section is negligible.

\subsection{$\alpha$ emission from the breakup of \textsuperscript{7}Li}
\label{sec:results7li}
\begin{figure}[!htb]
\centering
\includegraphics[width=0.85\linewidth] {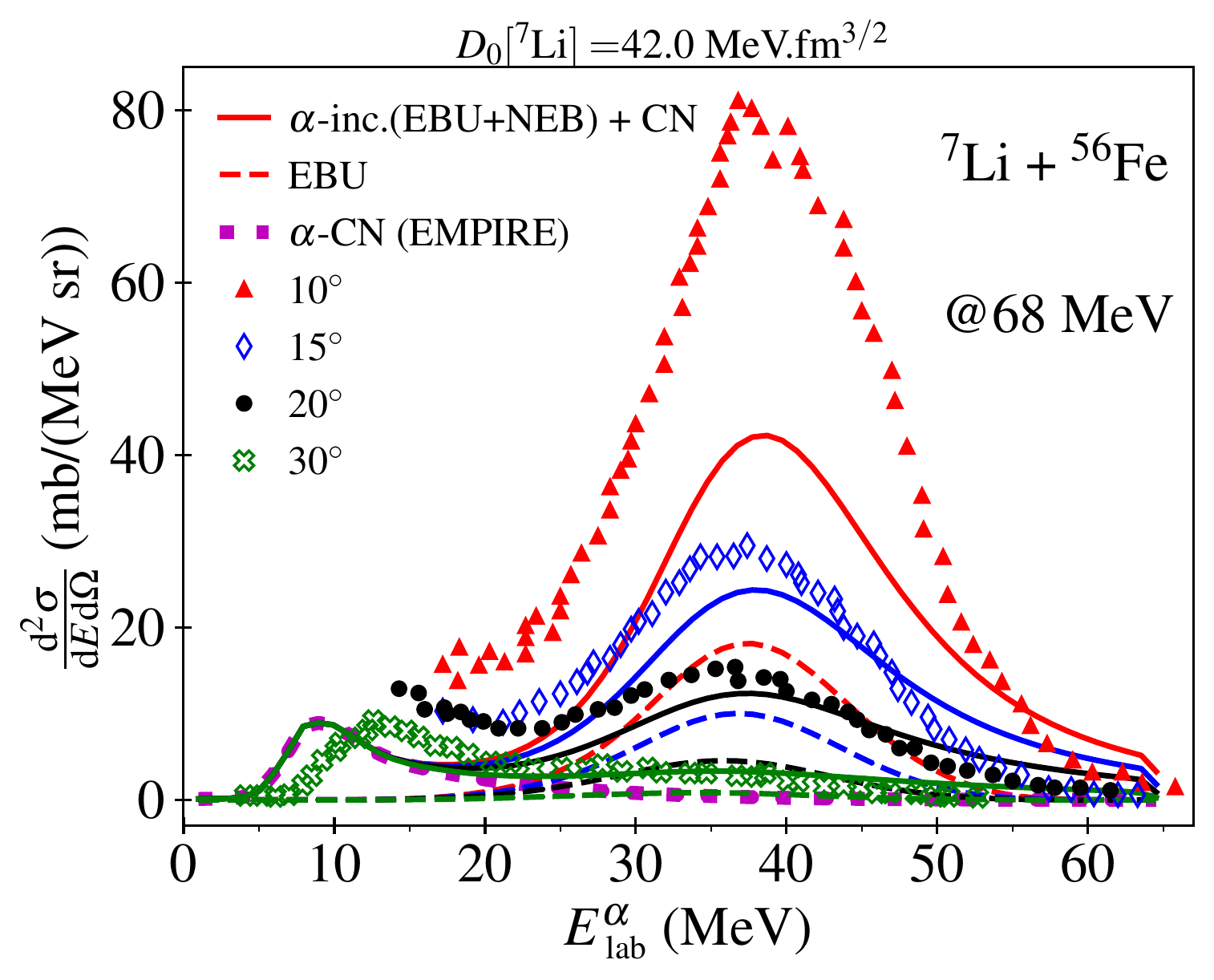}
\vspace{-0.5cm}
\caption{Double differential cross section of $\alpha$ emission from $^7$Li  colliding with $^{56}$Fe at a laboratory energy of 68 MeV for different angles. 
The experimental data are taken from Ref.~\cite{badran2001}.}
\label{fig:7Li-56Fe}
\end{figure}

\begin{figure}[!]
\centering 
\includegraphics[width=0.9\linewidth] {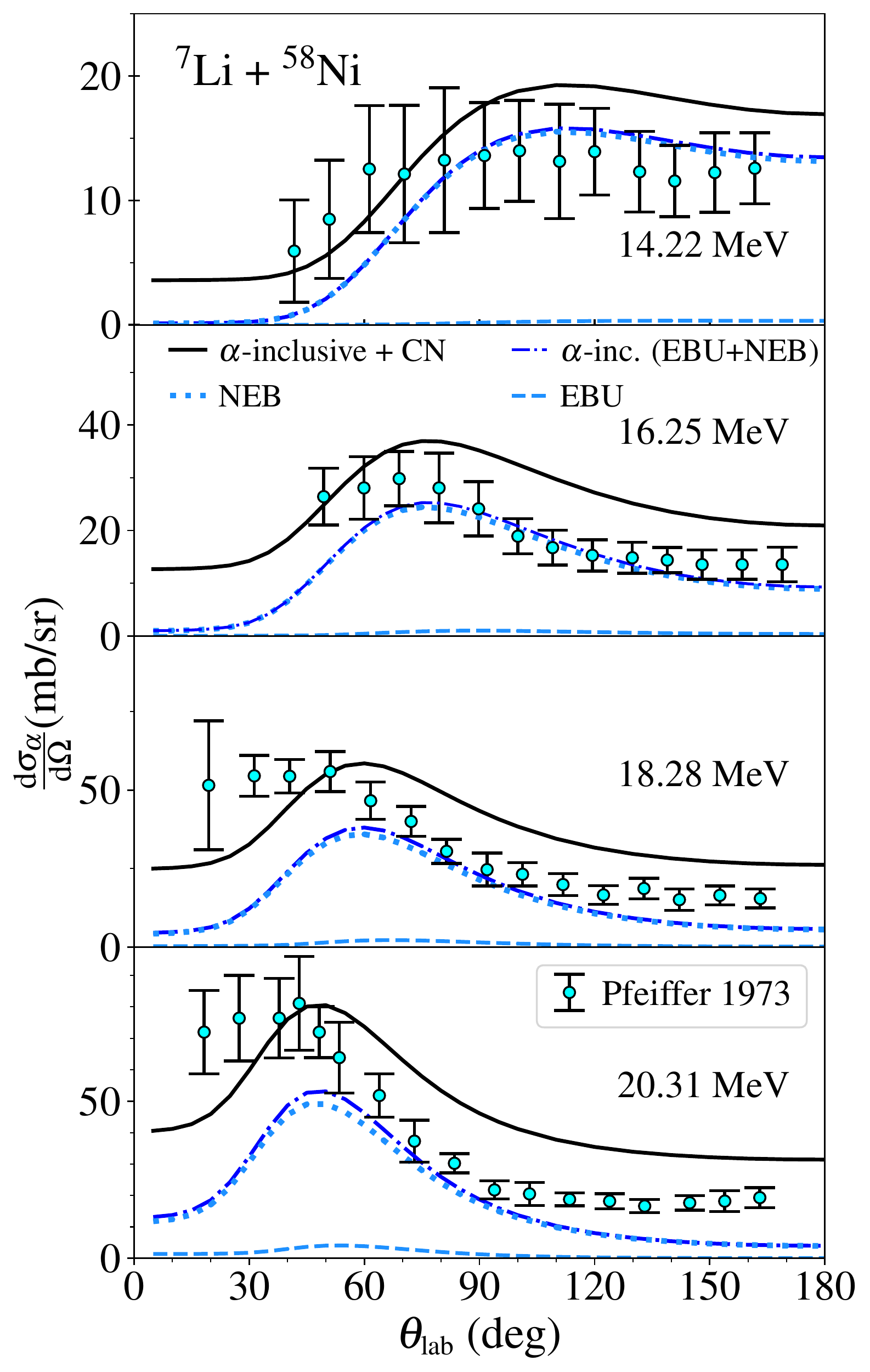}
\vspace{-0.5cm}
 \caption{$\alpha$ angular distribution in the collision of $^7$Li on a target of $^{58}$Ni compared to the experimental data taken from Ref.~\cite{pfeiffer1973}. The large contribution from the compound nucleus was calculated with the EMPIRE code and was assumed to be isotropic.
 \label{fig:7Li-58Ni}
 }
\end{figure}

We have modeled the structure of the nuclide $^7$Li as an $\alpha$+triton system and calculated inclusive $\alpha$ emission cross sections. The triton distorted wave functions were obtained with the Becchetti-Greenlees optical potential~\cite{becchetti1969}. Calculations were performed at 68 MeV $^7$Li incident energy on a $^{56}$Fe target and differential $\alpha$ emission spectra were compared with experimental data from Badran {\it et al.}  Ref.~\cite{badran2001}. {}{The value of $D_0=42.0 \textrm{ MeV.fm}^{3/2}$ is adjusted based on the double differential distribution shown in Fig.~\ref{fig:7Li-56Fe} and on the angular distribution of Fig.~\ref{fig:7Li-58Ni}. The $D_{0}$ value is chosen to best reproduce, simultaneously, most of the reaction data.}  The bumps presented in the experimental data in the low energy region ({$\approx~\text{13 MeV}$}) are associated with compound nuclear evaporation. One can see that at large scattering angles, the compound nucleus emission dominates at low energies.

Figure~\ref{fig:7Li-58Ni} shows inclusive $\alpha$ angular distributions from the reaction $^{58}$Ni($^7$Li,$\alpha$)X for different projectile energies. For this case, our calculations are again consistent with the trend of the experimental data. The large contribution from the compound nucleus component results in an overestimation of the reaction data. For comparison, we show in solid lines the inclusive breakup angular distribution summed with the compound nucleus contribution. 

Before closing we compare the $\alpha$ emission cross sections from the $^6$Li and $^7$Li induced reactions on $^{58}$Ni, shown respectively in Figs.~\ref{fig:6Li-58Ni} and~\ref{fig:7Li-58Ni}. The similar entrance energies of the two cases allow a comparison of $\alpha$ production from the different Lithium isotopes. 
The angular distributions are similar and are concentrated at backward angles at low incident energy. With increasing energy, a peak develops at more forward angles. The larger magnitude of the peak for the $^6$Li reaction would seem to be associated with the smaller binding energy of $^6$Li when compared to $^7$Li.

\begin{figure}[!]
\centering 
\includegraphics[width=0.9\linewidth] {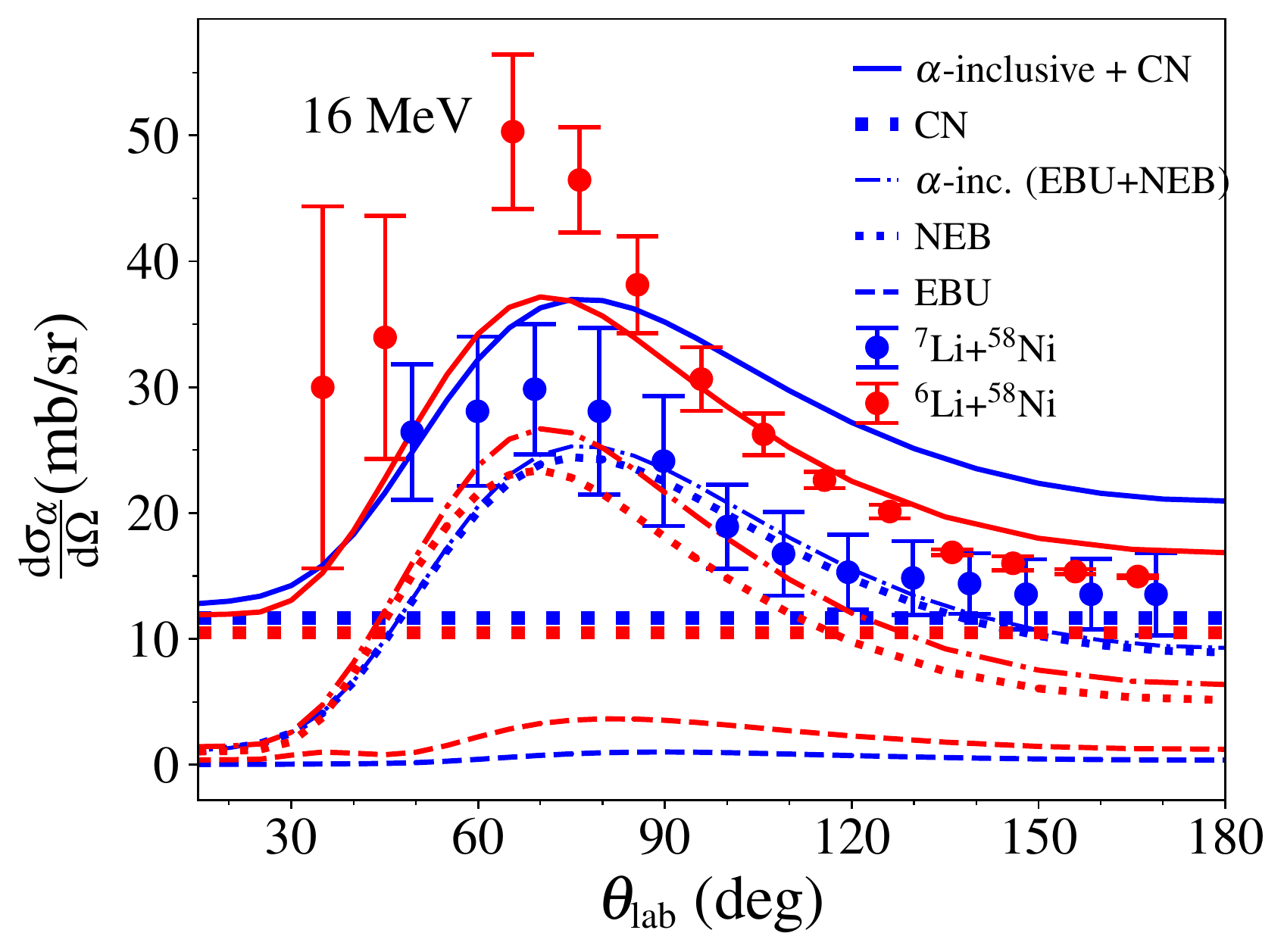}
\vspace{-0.5cm}
 \caption{Comparison of $\alpha$ angular distributions from the collisions of $^6$Li and $^7$Li with a target of $^{58}$Ni at 16 MeV. The experimental data are taken from Ref.~\cite{pfeiffer1973}.  
 \label{fig:6Li7Li-58Ni}
 }
\end{figure}

To facilitate this comparison, in Fig.~\ref{fig:6Li7Li-58Ni}, we plot the inclusive alpha emission angular distributions of both the $^6$Li and $^7$Li induced reactions at an incident energy of 16 MeV (Figs.~\ref{fig:6Li-58Ni} and~\ref{fig:7Li-58Ni}). Our complete calculations (the sum of CN and inclusive modes) show similar values for both projectiles. The enhancement of the $^6$Li experimental data commonly associated with the impact of the reduced binding energy on the elastic breakup component is not apparent in our calculations. As one can see in Fig.~\ref{fig:6Li7Li-58Ni}, the  nonelastic reaction mode dominates the angular distribution and furnishes similar distributions in shape and magnitude for the two projectiles. The contributions from the compound nucleus are also very similar in the two cases. As previously mentioned, alpha emission from the elastic breakup of $^6$Li is somewhat larger than that of $^7$Li. However, the difference is much smaller than that observed in the experimental data, which cannot be explained by our calculations.

\section{Summary and conclusions}\label{sec:conclusion}

Emission of $\alpha$ particles from breakup reactions of weakly bound nuclei are studied within the post-form IAV DWBA formalism. We have compared inclusive cross section calculations composed of elastic and nonelastic components to available experimental data. 
The distorted wave functions in the incident channel are obtained with the S\~ao Paulo optical potential. For simplicity, we have performed a Wood-Saxon parameterization of the SPP with a modified  diffuseness. The default value of the diffuseness parameter provided by the SPP leads to a too strongly absorptive interaction as a consequence of the fact that the SPP was developed for collisions involving heavier nuclei. 
The scattering wave functions of the unobserved/absorbed fragment $x$ and the spectator particle $b$ in the final states were generated using standard optical potentials with Woods-Saxon shapes. 
We have calculated total cross sections, energy spectra and double differential distributions of $\alpha$ particle emission. 
In addition to inclusive breakup, the compound nucleus contribution was computed using the EMPIRE code and included in the relevant cross sections and angular distributions. 
%

{}{
The angular distributions for $^6$Li and $^7$Li show a general trend of an under- overestimation of the data at small and large angles, respectively. A part of this trend could be due to the assumption of isotropic compound nuclear emission. At small angles, our calculations point to the need of inclusion of other components such as transfer to bound states. In the future we also intend to employ the CDCC method in comparison to our elastic analyses, due to its success in describing differential cross sections at small angles~\cite{duan2020}.
}

It is well known that elastic breakup does not reproduce the total strength of inclusive emission from the breakup reactions of weakly bound light nuclei.  In the case of $^6$Li dissociation on $^{90}$Zr, we show
that nonelastic breakup makes the dominant contribution to the $\alpha$ emission spectrum and to the total alpha emission cross section, except for large angles where the cross section is dominated by the CN emissions. 
In addition, our results indicate that the emitted alphas originate from the projectile through a direct reaction process.   
We found that our calculations are in good agreement with the experimental alpha particle emission cross sections for the isotopes $^{6 ,7}$Li. The case of $^6$He deserves further study both theoretically and experimentally. Although our calculations are in relatively good agreement with the experimental alpha emission spectra in this case, the differential cross section data is more forward peaked (especially at higher energies) when compared to our model predictions. Although we found good agreement of our calculations to the experimental data for distributions with $\theta>30^\circ$ in the case of the $^6$He+$^{64}$Zn reaction, the mechanism for particle emission at small angles is still in question. The angular distributions for $\alpha$ emission from the $^6$He+$^{120}$Sn reaction was also difficult to reproduce within our approach. The calculations presented here lie mostly below the experimental data for the angular distributions but follow well the shape of the double differential cross section.  In the two-body projectile model, the appropriate binding energy is taken as -1.6 MeV. We adjusted the zero range constant  $D_0$ based on this value. In addition, for the purpose of comparison, we have also calculated $^6$He angular distributions and spectra using the experimental two-neutron separation energy. 
One of the difficulties of our analysis is the description of the dineutron. The approach given here represents our first attempt to study such a reaction channel and will be better explored/improved in future studies. From the theoretical point of view, a complete four-body model could help to clarify questions about the reaction mechanisms of $^6$He, a very difficult problem to be handled at the moment. We would also like to mention the need for more experimental data for reactions of this kind.
Although one might expect the breakup mechanism to dominate the reactions studied here, the underestimated alpha production spectra from $^6$He breakup indicates that other processes should be explored.

We believe that further data at or near the Coulomb barrier are needed to better understand the competition between the final channels contributing to $\alpha$ production, especially in the case of the unstable nuclide $^6$He. %
In the future, we plan to improve our  description by extending the model to take into account more aspects of three-body breakup appropriate for the description of this nucleus.

%
\section{Acknowledgements}
   The authors acknowledge support from the S\~ao Paulo Research Foundation (FAPESP) under the grants 2017/05660-0 (BVC),  2016/07398-8 and 2017/13693-5 (EVC), and 2019/07767-1 (LAS). 
   BVC acknowledges grant 306433/2017-6 of the CNPq.  All the authors acknowledge the support of the INCT-FNA project 464898/2014-5. This work is performed in part under the auspices of the U.S. Department of Energy by Lawrence Livermore National Laboratory under Contract DE-AC52-07NA27344 with partial support from LDRD project 19-ERD-017.

\bibliographystyle{apsrev4-2}
\bibliography{bib.bib}

\end{document}